\documentclass{aip-cp}

\usepackage[numbers]{natbib}
\usepackage{rotating}
\usepackage{graphicx}

\begin{document}
\newcommand{\aap}{A\&A}
\newcommand{\apj}{ApJ}
\newcommand{\aj}{AJ}
\newcommand{\apjl}{ApJL}
\newcommand{\mnras}{MNRAS}
\newcommand{\pasa}{PASA}
\newcommand{\prd}{Phys.Rev.D}
\newcommand\lppr{\lower.5ex\hbox{$\; \buildrel < \over \sim \;$}}
\newcommand\gppr{\lower.5ex\hbox{$\; \buildrel > \over \sim \;$}}

\title{Gamma-Rays from Non-Blazar AGN}
    
\author[aff1]{Frank M. Rieger}

\affil[aff1]{Zentrum f\"ur Astronomie (ZAH), Institut f\"ur Theoretische Astrophysik, 
Universit\"at Heidelberg, Philosophenweg 12, 69120 Heidelberg,\\ \& \\ Max-Planck-Institut 
f\"ur Kernphysik, P.O. Box 103980, 69029 Heidelberg, Germany}
\eaddress{frank.rieger@mpi-hd.mpg.de}

\maketitle

\begin{abstract}
Non-blazar Active Galactic Nuclei (AGN) have emerged as a new $\gamma$-ray emitting source class on 
the extragalactic sky and started to deepen our understanding of the physical processes and the nature of 
AGN in general. The detection of Narrow Line Seyfert 1 galaxies in the Fermi-LAT energy regime, for example, 
offers important information for our understanding of jet formation and radio-loudness. Radio galaxies, on 
the other hand, have become particularly interesting at high (HE) and very high (VHE) gamma-ray energies. 
With their jets not directly pointing towards us (i.e. "misaligned"), they offer a unique tool to probe into the
nature of the fundamental (and often "hidden") physical processes in AGN. 
This review highlights and discusses some of the observational and theoretical progress achieved in the 
gamma-ray regime during recent years, including the evidence for unexpected spectral hardening in 
Centaurus A and extreme short-term variability as seen in IC~310 and M87. 
\end{abstract}

\section{INTRODUCTION}
Within the last couple of years, the extragalactic sky has become populated by gamma-ray emitting
sources.  The Fermi-LAT Third AGN catalog (3LAC) now list about 1600 HE (high energy, $>100$ MeV) 
sources \citep{2015ApJ...810...14A}, while at VHE (very high energy, $>100$ GeV) energies the 
detection of about 70 AGN is currently summarised in the TeVcat catalog\footnote{http://tevcat.uchicago.edu}.
Most of these sources are of the blazar type, i.e. radio-loud AGN in which the jet is thought to be inclined 
at small viewing angles $i$ to the line of sight. This goes along with substantial Doppler-boosting of their 
intrinsic jet emission, $S(\nu)=D^{\alpha}S'(\nu')$ where $D=1/[\gamma_b(1-\beta_b\cos i)]$ is the Doppler
factor, $\gamma_b$ the jet bulk Lorentz factor and $\alpha\geq 2$, which favours their detection on the sky 
by current gamma-ray instruments.\\ 
Non-blazar AGN, such as Radio Galaxies or Narrow Line Seyfert 1 galaxies, are much less occurrent, but 
have in the meantime solidly emerged as new gamma-ray emitting source classes as well (see below):\\ 
{\it Radio Galaxies (RG)} have been conventionally classified as radio-loud (R$>$10), lower-luminous 
($M_v>$-23) AGN, hosted by elliptical galaxies and often revealing broad or narrow emission lines in their 
spectra (shortened BLRG and NLRG). Morphologically, radio galaxies have also been categorised since long 
into Fanaroff-Riley I and II sources (FR I, FR II) \citep{1974MNRAS.167P..31F}, the former (FR I) class 
comprising lower luminosity ($<2\times10^{26}$ W/Hz at 178 MHz), edge-darkened sources and the latter 
one (FR II) high luminous ($>2\times10^{26}$ W/Hz at 178 MHz), edge-brightened sources where the radio 
lobes are dominated by bright hot spots. Various considerations suggest that the central engine in FR II might 
be accreting in a "standard" (geometrically thin, optical thick) mode, while FR I sources are probably supported 
by a radiatively inefficient accretion flow (RIAF).\\
{\it Narrow Line Seyfert 1 (NLSy1)} galaxies \citep{2008RMxAC..32...86K}, on the other hand, are optically 
lower luminosity ($M_v > -23$) AGN which are typically found in spiral/disk galaxies, spectroscopically 
revealing strong emission lines, in particular narrow (FWHM$_{H\beta} <2000$ km/s) optical Balmer (H$\beta$) 
lines from the broad line regime (BLR) and strong Fe II emission (bump).  At X-ray energies, NLSy1 are found 
to exhibit significant variability, a steep intrinsic spectrum and a relatively high luminosity. In view of their smaller 
black holes masses ($10^{6-8} M_{\odot}$), as estimated from reverberation techniques, this would seem to 
imply accretion rates close to the Eddington one. NLSy 1 are mostly radio-quiet ($R<10$), with only a small
fraction ($< 10\%$) being radio-loud.\\

\section{CURRENT STATUS AT GAMMA-RAY ENERGIES}
\begin{figure}[h]
  \centerline{\includegraphics[width=350pt]{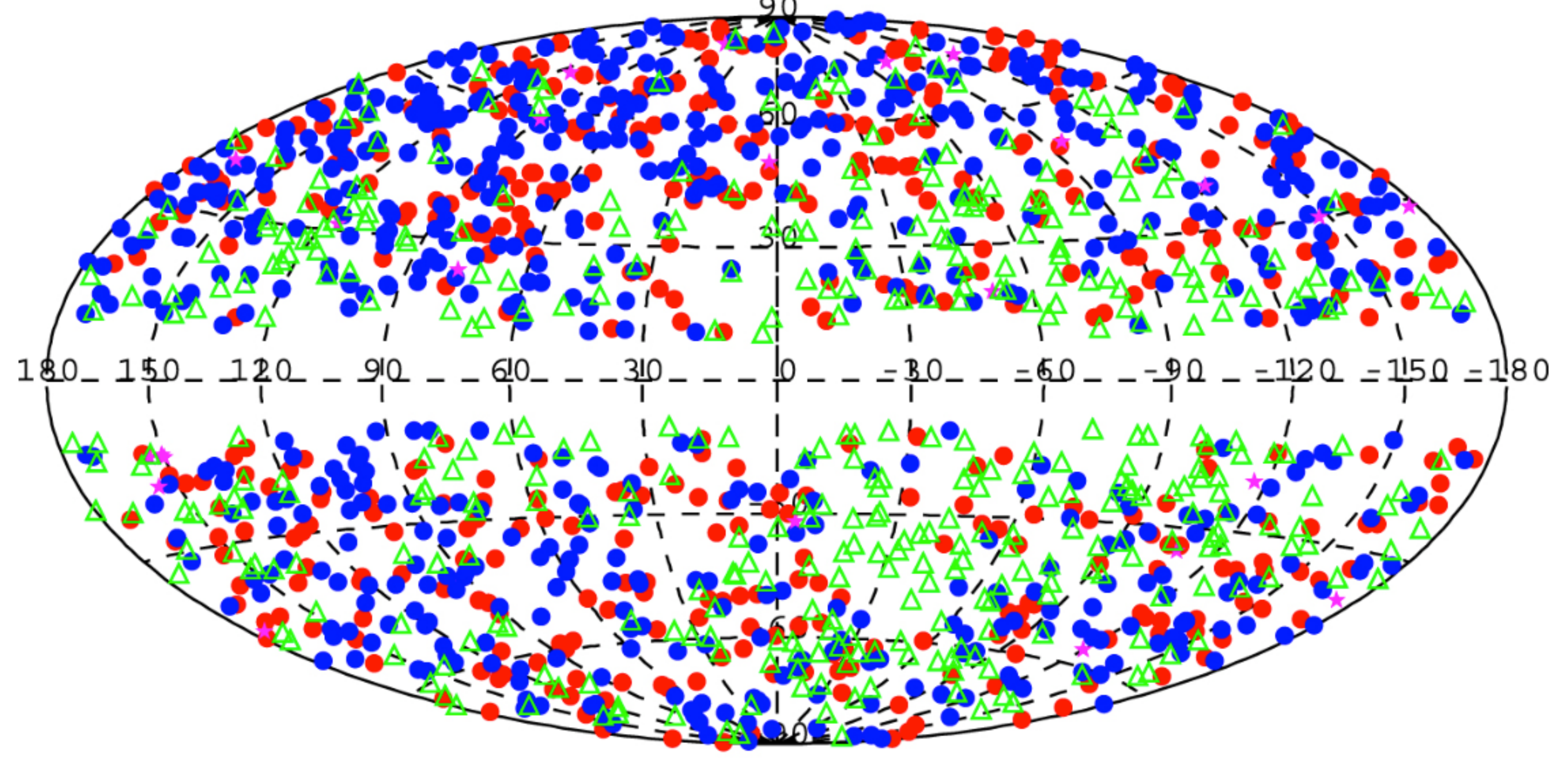}}
  \caption{Locations of AGN sources at HE gamma-rays (clean sample) as reported in the Fermi-LAT Third 
  AGN Catalog. Red circles classify: FSRQs; blue circles: BL Lacs; green triangles: blazars of uncertain 
  type; magenta stars: other AGN including radio galaxies. From Ref.~\citep{2015ApJ...810...14A}, $\copyright$ 
  AAS, reproduced with permission.}\label{3LAT_AGN}
\end{figure}

At HE gamma-rays, Fermi-LAT has reported the detection of 1591 (1444 in the clean sample) AGN (with TS 
values$>$25) at $|b| >10^o$ in its 3 LAC catalog (published in 2015) based on 48 months of data (see 
Fig.~\ref{3LAT_AGN}), out of which 467 have been classified as Flat Spectrum Radio Quasars (FSRQs), 632 
as BL Lacs, 460 as blazar candidates of uncertain type and 32 as non-blazar AGN (then including 5 NLSy1). 
In terms of numbers, non-blazar AGN thus make out only a small fraction ($\lppr 2\%$) of all extragalactic 
gamma-ray sources. Despite this, they have started to strongly impact on the field by offering unique insights 
into the physics and astrophysics of accreting supermassive black hole systems.

\subsection{NLSy 1} 
While by now the detection of nine NLSy1\footnote{1H 0323+342, SBS 0846+513, PMN J0948+0022, PKS 
1502+036, PKS 2004-447, FBQS J1644+2619, B3 1441+476, NVSS J124634+023808, 4C +04.42} has been 
reported  at HE gamma-rays, none of them has yet been seen at VHE energies. In the Fermi-LAT energy range, 
these NLSy1 sources appear relatively bright given their known redshifts (up to $z\simeq 0.58$), implying apparent 
isotropic luminosities of $L_{\rm GeV} \sim 10^{44-48}$ erg/s \citep{2015MNRAS.446.2456D}. Current models 
suggest, though, that the non-thermal emission is non-isotropic and arises in a relativistic jet. There is in fact 
increasing evidence for the presence of relativistic outflows in some of these radio-loud NLSy1, such as a 
one-sided core-jet radio structure on pc-scales, superluminal ($\beta_{app}\sim10$) motion of components 
(SBS 0846+513, z=0.583) and intense "blazar-like" variability (optical intraday, HE flaring) suggestive of beaming, 
e.g. \citep{2012MNRAS.426..317D}.\\ 
The spectral energy distributions (SED) of the detected NLSy1 appears to be double-humped shaped, suggestive 
of a synchrotron-inverse Compton origin, where the HE gamma-rays probably arise due to external Compton 
scattering of seed photons from the BLR and/or from a dust-torus. The non-detection at VHE energies may then 
be due to an intrinsic limit or related to the BLR or EBL (Extragalactic Background Light) absorption of their TeV 
emission. The "unusual" presence of jets in an AGN class believed to be hosted by spiral galaxies provides 
important information for our common understanding of jet launching in astrophysical sources, e.g., with respect 
to a putative black hole mass threshold, favourable accretion states and black hole spin (and merger history) 
requirements and unification scenarios \citep{2009ApJ...699..976A,2015A&A...575A..13F}. 

\subsection{Radio Galaxies}
\begin{figure}[h]
  \centerline{\includegraphics[width=350pt]{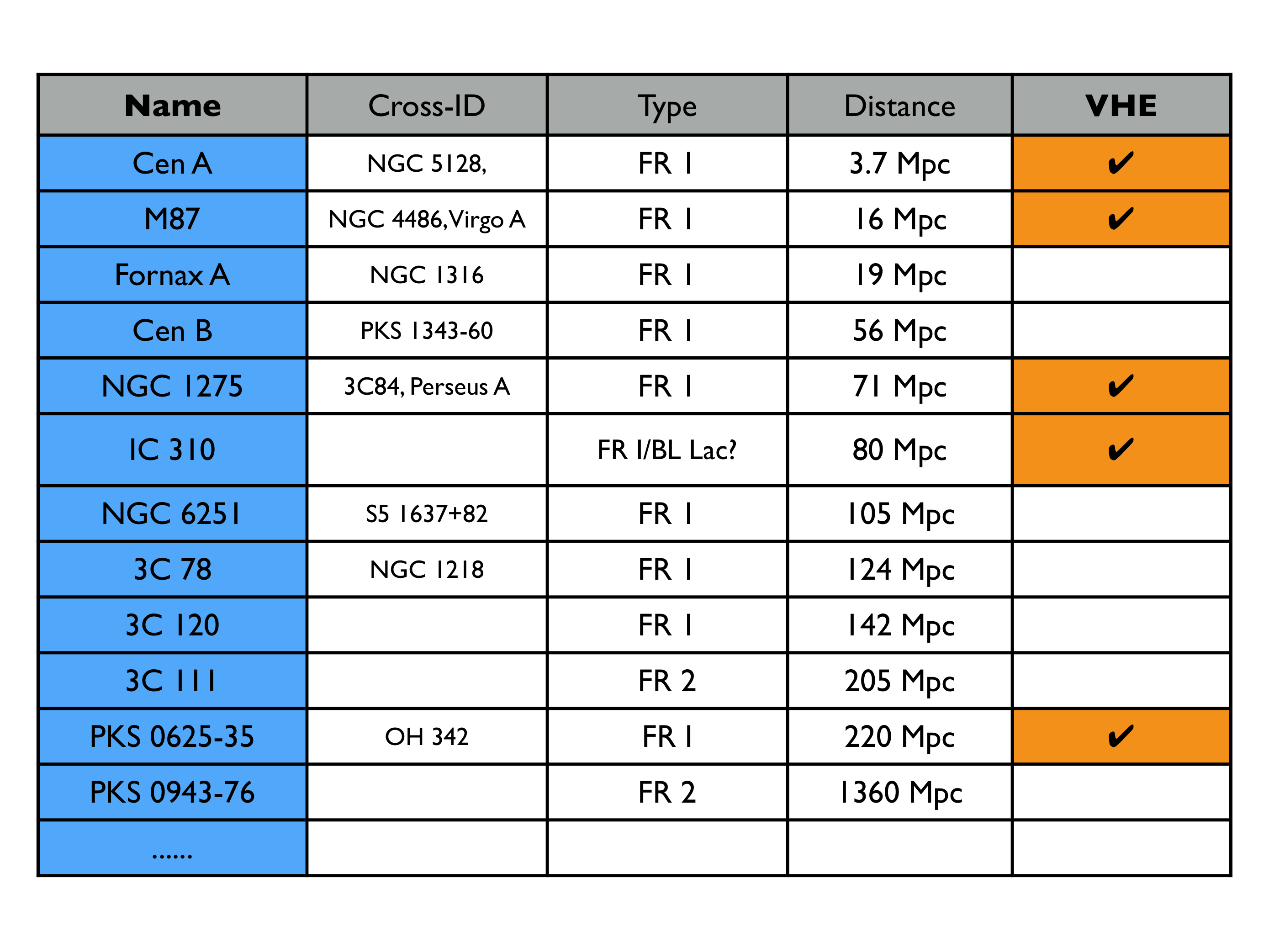}}
  \caption{Selection of radio galaxies detected at gamma-ray energies (Cross-IDs give their alternative identifications). 
  Out of the Fermi-LAT detected HE sources, only five have so far been seen at TeV energies.}\label{rg_list}
\end{figure}

Out of the tenish radio galaxies detected at HE gamma-rays by Fermi-LAT, cf. also Ref.~\citep{Narek16,Angioni16}, 
five have so far been seen at TeV energies (see Fig.~\ref{rg_list}). M87 ($d\sim 16$ Mpc) has been the first detected in 
VHE gamma-rays, while PKS 0625-35 ($z=0.055$) has been the latest addition. At $d\sim 4$ Mpc Centaurus~A (Cen A) 
is the nearest one, while PKS 0625-35 at $d\sim 200$ Mpc is farthest away.\\
\noindent
The jets in radio galaxies are commonly thought to be substantially misaligned ($i>10^o$), suggesting rather modest 
Doppler factors ($D\leq$ a few) and thus moderate Doppler boosting only. In the simplest AGN unification scheme, 
FR~I type sources are taken to resemble BL Lac objects (a blazar subclass) viewed at large viewing angles. The 
double-humped shaped SEDs of BL Lac-type sources have for long been satisfactorily fitted by 
(single-zone) leptonic synchrotron-self Compton (SSC) models with high Doppler factors, assuming the second hump 
to be related to inverse Compton up-scattering of synchrotron photons by the electrons themselves. Earlier application 
to radio galaxies suggested that their (then available) nuclear SED could be fitted by similar (SSC-jet) processes yet 
with small Doppler factors ("misaligned BL Lacs") \citep{2001MNRAS.324L..33C}. If this would indeed apply, the 
emission of only a few radio galaxies are expected to reach into the Fermi-LAT GeV regime (as seems indeed to be 
the case), and little sources are expected to show up at TeV energies. Yet, despite being relatively small in number, 
RG have turned out particularly interesting by offering unique insights into some of the fundamental (and often hidden) 
non-thermal processes in gamma-ray emitting AGN. In the following a brief description of the individual source status
is given.

\subsubsection{PKS 0625-354}
PKS 0625-354 was discovered as a VHE emitter above 250 GeV (at a level of $\sim6\sigma$ in 5.5 h of data) in 
2012 by H.E.S.S. \citep{2015arXiv150906851D}. Its VHE spectrum extends to $\sim 2$ TeV and seems rather 
steep (compatible with a power law of photon index $\sim -2.8\pm0.5$). The associated apparent isotropic VHE
luminosity is moderate and of the order of $L_{\rm VHE} \sim 5 \times 10^{42}$ erg/s.
PKS 0625-354 is an AGN at a distance of about $d\sim 220$ Mpc with a black hole mass (inferred from the
M-bulge relation) of the order of $M_{\rm BH}\sim 10^9 M_{\odot}$ \citep{2014MNRAS.440..269M} and 
probably accreting in an inefficient mode. It is known as a low excitation line radio-loud object, but there is 
uncertainty as to whether it resembles more a BL Lac or FR I-RG type source, e.g. \citep{2004MNRAS.347..771W}.
Recent TANAMI observations provide evidence for a one-sided parsec-scale jet and superluminal motion 
with $\beta_{\rm app} \sim 3$ \citep{2013arXiv1301.4384M}, rather supporting a BL Lac classification along 
with non-modest Doppler boosting. Its classical "radio galaxy" status may thus have to be re-considered.  
No evidence for significant VHE variability has been found in the H.E.S.S. data set, although hints for a HE 
variability on a timescale $\sim 100$ d might be inferred from its Fermi-LAT light curve, cf. also 
\citep{2015ApJ...798...74F}. Given its unclear classification and the limited VHE data set, the current situation 
seems physics-wise not yet constraining enough to draw robust conclusions on the origin of its non-thermal emission 
much beyond single zone SSC-type considerations \citep{2015ApJ...798...74F}. To some extent PKS 0625-354 may 
thus remind one of the early stages of VHE blazar research, with future surprises not excluded.

\subsubsection{NGC 1275} 
NGC 1275, also known as Perseus A, is the central radio galaxy of the Perseus cluster of galaxies at a distance 
of $\sim 70$ Mpc. The source was detected as VHE emitter above hundred GeV by MAGIC with a flux level of 
$\sim3\%$ of the Crab Nebula, and studied in two observational campaign (the first during 10/2009-02/2010 with 
$\sim46$h at a level of $6.1\sigma$; the second during 08/2010-02/2011 with $\sim54$ h at a level of $6.6\sigma$) 
\citep{2012A&A...539L...2A, 2014A&A...564A...5A}. Its VHE spectrum, when characterised by 
a single power law alone, is very steep with photon index of $\sim-4.1$. No signal has been seen above 650 GeV. 
When HE (Fermi-LAT) and VHE data are combined, the gamma-ray spectrum appears compatible with either a 
log-parabola or a power-law with a sub-exponential cut-off, suggestive of a common origin and of a peak or cut-off 
around several GeV. While at VHE energies only hints for month-type variability could be established, the source 
is known to show significant HE variability on timescales of a few days \citep{2011MNRAS.413.2785B}, suggesting 
that the gamma-ray emission originates in a (possibly, single) compact zone. 
NGC 1275 hosts a supermassive black hole of mass $\sim3 \times 10^8 M_{\odot}$ \citep{2005MNRAS.359..755W} 
and shows a pc-scale radio jet orientated at $i\sim 30-45^o$ \citep{1994ApJ...430L..45W, 2016arXiv160904017F}.
A one-zone SSC interpretation of its SED, assuming the sub-pc scale jet to be misaligned ($i\sim20-30^o$) (i.e. the 
"classical" misaligned BL Lac scenario), appears possible, though some tension may arise with the inferred jet 
inclination on pc-scales. This could probably be alleviated if e.g. the emitting component follows a non-straight 
trajectory that relaxes with distances, or if the jet is structured (spine-shear) allowing for a more complex inverse
Compton interplay \citep{2014MNRAS.443.1224T}.

\subsubsection{Centaurus~A}
\begin{figure}[h]
  \centerline{\includegraphics[width=320pt]{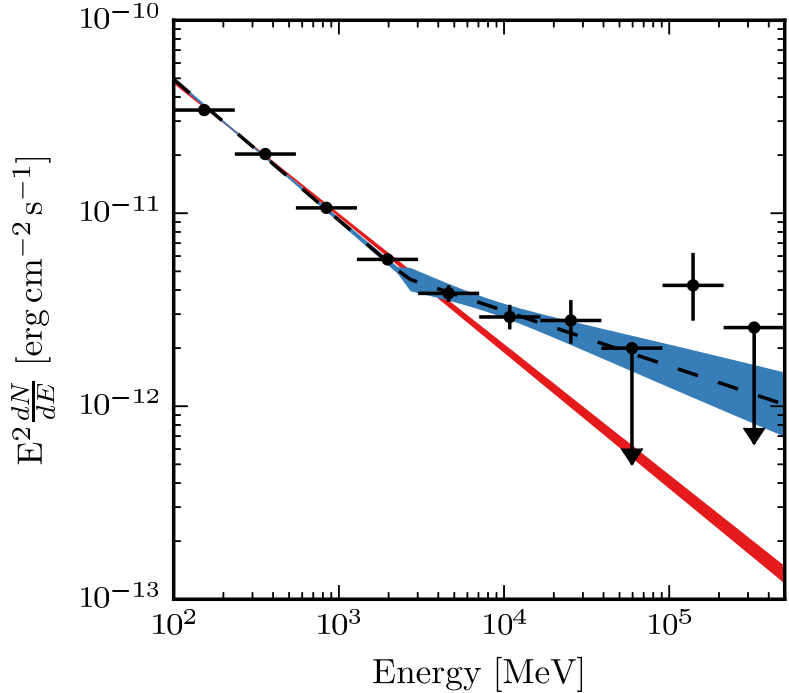}}
  \caption{The high-energy gamma-ray spectrum of Cen A above 100 MeV based on $\sim 7.5$ year of 
  Fermi-LAT data. The spectrum shows an unexpected spectral hardening at a few GeV, where its index 
  changes by $\sim 0.5$ (assuming a broken power law), see also Refs.~\citep{2013ApJ...770L...6S,
  2016arXiv160305469B}. This feature is most naturally interpreted as a physically distinct emission 
  component that emerges towards highest energies and allows to smoothly connect the HE and VHE 
  regimes. Figure courtesy of J. Graham.}
  \label{CenA_SED}
\end{figure}

\begin{figure}[h]
  \centerline{\includegraphics[width=450pt]{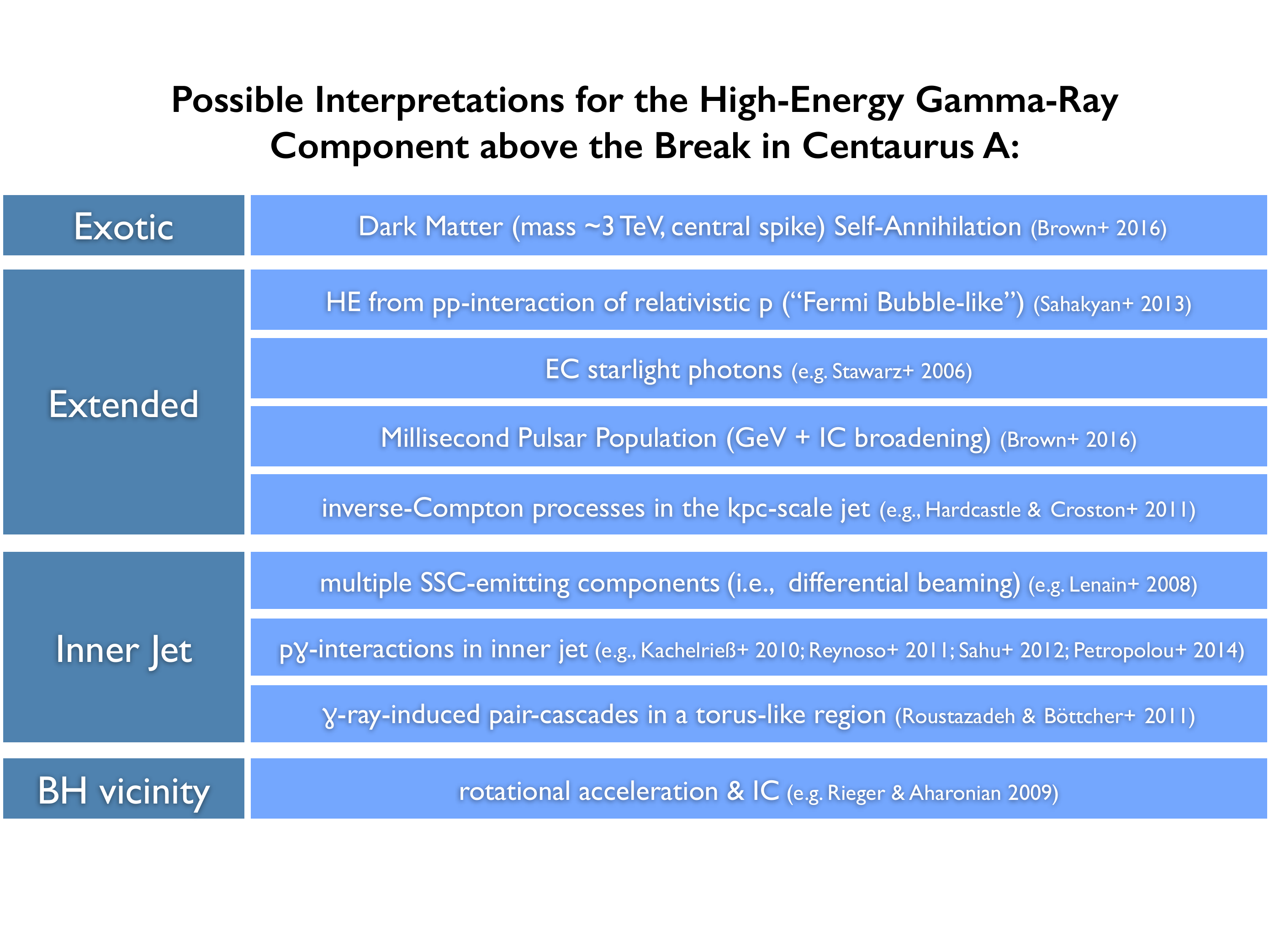}}
  \caption{Possible scenarios for the nature of the high energy gamma-ray component (above the break) 
  seen in Cen~A.}\label{interpretations}
\end{figure}

Being the nearest AGN on the sky at a distance of $d\simeq 3.7$ Mpc, Centaurus A (Cen~A) is one of the best
studied extragalactic objects. It hosts a black hole of mass $(0.5-1) \times 10^8 M_{\odot}$ and shows (under 
the proviso of adopting a quasar SED template) an estimated bolometric luminosity $L_{\rm bol} \sim 10^{43}$ 
erg/s much less than the expected Eddington luminosity $L_{\rm Edd}$, suggesting that accretion in its inner 
disk part might occur in a radiatively inefficient mode, cf. also \citep{2016ApJ...819..150F}. 
Observations at radio frequencies have revealed a peculiar morphology with evidence for a sub-pc scale jet 
and counter-jet, a one-sided kpc jet, two radio lobes and extended diffusive emission. VLBI studies suggest 
that Cen~A is a "non-blazar" source, its inner jet probably being substantially inclined (e.g., $i\sim 12-45^o$ 
based on TANAMI jet-counter jet flux ratio measurements, under the proviso of intrinsically symmetric jets) 
and characterized by moderate (radio) bulk flow speeds $u_j  < 0.5$ c only, e.g. \citep{2001AJ....122.1697T,
2014A&A...569A.115M}.\\  
The detection of VHE gamma-rays (at a level of $5\sigma$) from the core region of Cen~A has been reported 
by H.E.S.S. based on more than $100$h of data taken in 2004-2008 \citep{2009ApJ...695L..40A}. The VHE 
spectrum extends from 300 GeV up to $\sim5$ TeV and seems relatively hard (compatible with a single power 
law index $\Gamma \simeq -2.7\pm0.5$). The source is relatively weak at VHE energies with an equivalent 
apparent isotropic luminosity of $L(>250$ GeV$) = 2.6 \times 10^{39}$ erg/s. No significant VHE variability has 
been detected, so that an extended (within the angular resolution $\sim 0.1^o$ of H.E.S.S.) origin of the TeV 
emission cannot be simply discarded.\\
At HE, Fermi-LAT has detected gamma-ray emission from both the core (within $\sim 0.1^o$) and the giant 
lobes of Cen~A \citep{2010ApJ...719.1433A,2010Sci...328..725A,2012A&A...542A..19Y,2016arXiv160603053S}. 
Together with Fornax A \citep{2016ApJ...826....1A}, extended HE gamma-ray emission has thus by now been 
seen in two radio galaxies, allowing to explore the associated plasma dynamics, radiation fields and processes 
on large scales.\\
The core region of Cen~A was initially detected up to 10 GeV (at a level of $4\sigma$) in 10 month of data 
and the HE spectrum then appeared compatible to a single power law with $\Gamma -2.67\pm 0.1$. While 
this index is very close to what has been seen at VHE, it became soon apparent that a simple HE power-law 
extrapolation would under-predicted the fluxes measured at TeV energies. The HE and VHE data were taken 
non-simultaneously, but the absence of variability in both energy bands suggested that the discrepancy might
be continuing. An update using 4 year of Fermi-LAT data (allowing for spectral extension up to $\sim 50$ GeV),
then found intriguing evidence for an unusual spectral hardening at HE gamma-ray energies with a possible
break $E_b\sim 4$ GeV where (in a broken power law approximation) the photon index changes from about 
$-2.7$ (below $E_b$) to about $-2.1$ (above $E_b$), with associated apparent luminosities $L(>E_b) \sim 
10^{40}$ erg/s, $L(<E_b) \sim 10^{41}$ erg/s, respectively \citep{2013ApJ...770L...6S}. A recent update using
$7.5$ year of Fermi-LAT data (enabling spectral extension up to $\sim 150$ GeV) \citep{2016arXiv160305469B}, 
also reported in these Proceedings, is reinforcing this conclusion of spectral hardening at a $>5\sigma$ level, 
see Fig.~\ref{CenA_SED}. While in the AGN context spectral steepening at gamma-ray energies is familiar, a 
spectral hardening comes quite unexpected. This spectral feature is most naturally understood as revealing 
the emergence of an additional emission component that extends into the TeV regime and that is beyond the 
conventional (single and strongly Doppler-boosted) SSC-contribution which often seems to account for the 
SEDs in blazars. 
From a astrophysical points of view, this is particularly interesting and a variety of different (not mutually
exclusive) interpretations as to its true origin are encountered in the literature, see e.g. 
Fig.~\ref{interpretations}.  
These include (i) seemingly exotic suggestions such as the self-annihilation of dark matter particles of 
mass $\sim 3$ TeV in a putative central dark matter (DM) spike) \citep{2016arXiv160305469B}, (ii) extended 
scenarios based on e.g. the hadronic interaction of relativistic protons with ambient matter in the large-scale
 (halo-size) region ("Fermi-bubble like") \citep{2013ApJ...770L...6S}, the integrated gamma-ray contribution 
 from a supposed population of millisecond pulsars \citep{2016arXiv160305469B}, or (leptonic) inverse-Compton 
 processes with various photon fields (SSC, starlight, CMB, EBL) in the kpc-scale jet of Cen~A
 \citep{2006MNRAS.371.1705S,2011MNRAS.415..133H}, (iii) inner (pc-scale an below) jet models such a multiple 
(leptonic) SSC-emitting components moving at different angles to the line of sight \citep{2008A&A...478..111L}, 
photo-meson (p$\gamma$) interactions of ultra-high energy protons in strong photons fields (e.g. standard disk-type) 
on inner jet scales \citep{2010PASA...27..482K, 2012PhRvD..85d3012S,2014A&A...562A..12P} and elaborated 
lepto-hadronic variants thereof \citep{2011A&A...531A..30R,2016arXiv161000255C}, or $\gamma$-ray induced 
pair-cascades in a putative dusty torus-like region \citep{2011ApJ...728..134R}, last not least (iv) magnetospheric 
scenarios based on leptonic inverse Compton processes in an under-luminous ADAF-type environment 
\citep{2009A&A...506L..41R,2011IJMPD..20.1547R}.\\ 
All of these models come with some challenges, from e.g. an anomalously high dark matter concentration in 
DM scenarios, a high required jet power in hadronic models and poorly-known density profiles for pulsars to 
the deviation from equipartition and internal opacity constraints in leptonic models, the putative absence of a 
dusty torus in Cen~A or external opacity constraints for near black hole scenarios.\\ 
There is evidence for variability of the gamma-ray emission below the break $E_b$ on timescale of several 
months \citep{2016arXiv160305469B}, which seems compatible with a standard (e.g. one-zone SSC) jet origin 
of this emission. At higher energies (i.e., for the so-called 2nd component) no significant variability has yet 
been detected, though the limited Fermi-LAT statistics and the weakness of the source in the VHE regime 
are making it difficult to really probe into this. Given the limited angular resolution ($\sim 5$ kpc) of current 
gamma-ray instruments, uncertainties thus remain as to the true origin of the $\gamma$-ray emission above 
the break, and both extended and inner jet-related scenarios appear possible. We note that possible hints 
for variability would allow to disfavour the former ones, while the current (apparent) lack of variability does 
not necessarily provide compelling evidence against an inner jet-related origin if the jet is indeed sufficiently 
misaligned and Doppler-boosting effects are weak.\\ 
Apart from some circumstantial evidence in the case of the gamma-ray blazar Mkn~501, this is the first
time that spectral results provide strong evidence for the appearance of a physically distinct component
at $\gamma$-ray energies. While it seems that up to a few GeV the SED of Cen~A could in principle be 
reasonably well modelled by standard leptonic SSC processes assuming a "misaligned blazar" 
\citep{2001MNRAS.324L..33C,2010ApJ...719.1433A}, the true nature of this second component and its 
relation to the AGN still remains to be disclosed.

\subsubsection{IC 310}
\begin{figure}[h]
  \centerline{\includegraphics[width=320pt]{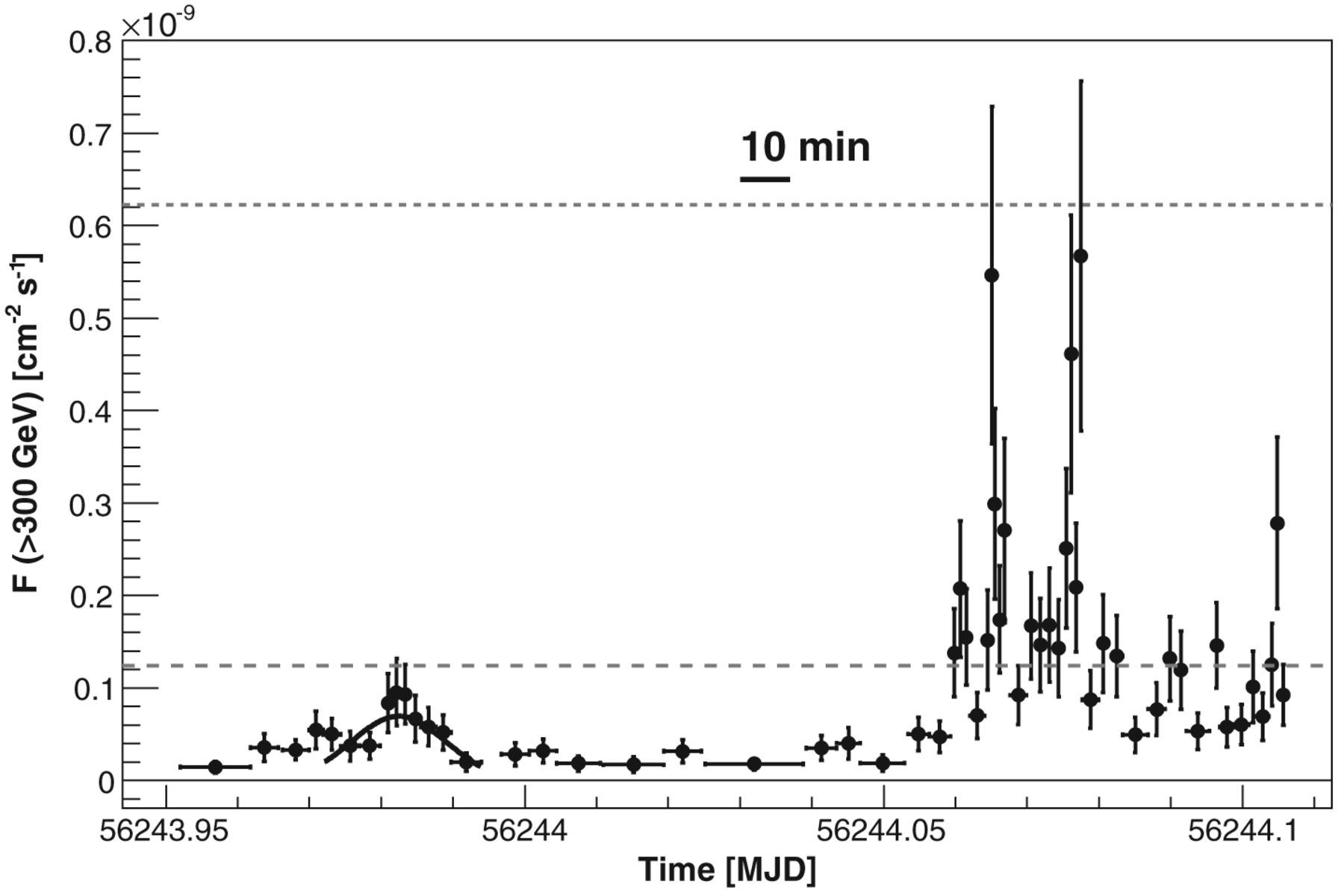}}
  \caption{VHE light curve of IC 310 above 300 GeV as observed with the MAGIC telescopes during November 
  12/13, 2012. Extreme VHE variability on (doubling) timescale much less then 10 min is apparent from the Figure. 
  The two gray lines indicate  flux levels of 1 and 5 Crab units, respectively. From Ref.~\citep{2014Sci...346.1080A}. 
  Reprinted with permission from AAAS.}\label{IC310}
\end{figure}

\begin{figure}[h]
  \centerline{\includegraphics[width=400pt]{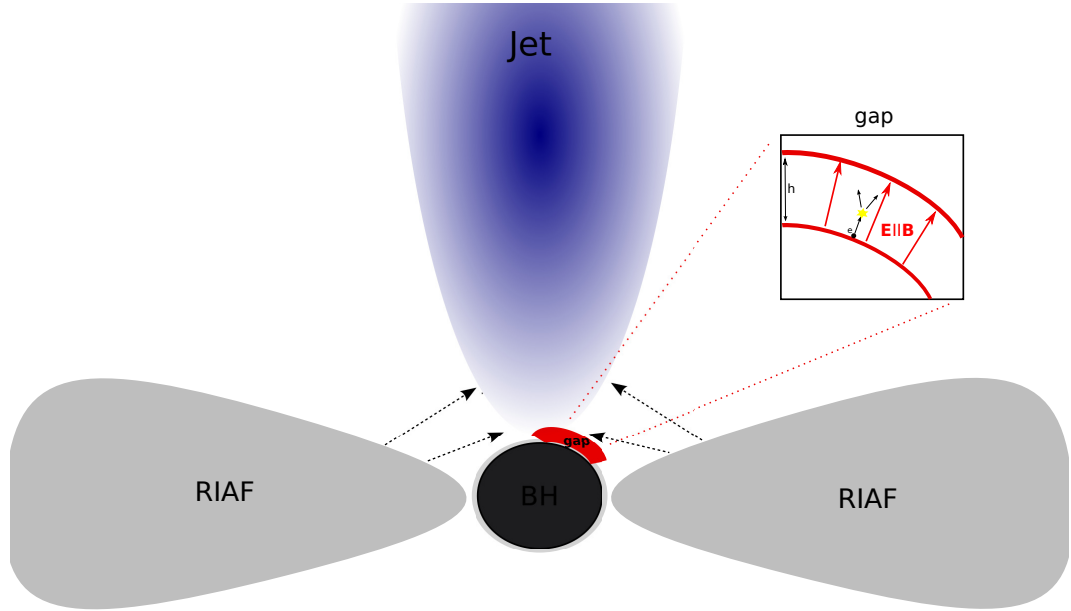}}
  \caption{Unscreened electric field regions ("vacuum gaps" where $\vec{E}\cdot\vec{B}\neq 0$) may form close to the 
  horizon in under-dense black hole environments. Charged seed electrons, injected by e.g. pair-creation processes in 
  the inner accretion flow (RIAF) into these regions, are then quickly accelerated along the fields to high energies and 
  can trigger an electromagnetic pair cascade, that is accompanied by highly variable VHE emission (due to curvature 
  radiation or inverse Compton processes) and that eventually ensures a charge supply high enough to establish the 
  formation of jet like features. A scenario of this type has been proposed to account for the extreme VHE variability in 
  IC~310 \citep{2014Sci...346.1080A}.}\label{gap}
\end{figure}

The Perseus Cluster radio galaxy {\it IC~310} at a distance of $d\sim 80$ Mpc has been detected at VHE 
energies by MAGIC in about of 21~h data (taken during 10/2009-02/2010)\citep{2010ApJ...723L.207A}.
The source, which is believed to host a black hole of mass $\simeq 3\times 10^8 M_{\odot}$ 
\citep{2014Sci...346.1080A}, has for a while been classified as a head-tail radio galaxy, but the apparent 
lack of jet bending along with recent indications for a one-sided pc-scale radio jet structure inclined at 
$i\lppr 38^o$ suggests that IC~310 may instead be a source at the borderline dividing low-luminosity 
radio galaxies and BL Lac objects \citep{2012A&A...538L...1K}. The spectrum at VHE energies, which has 
been measured up to $\sim 10$ TeV, is very hard and compatible with a single power law (there is no 
evidence for a break) of photon index $\Gamma \simeq -2.0$ \citep{2014A&A...563A..91A}. The VHE flux 
in the 2009-2010 campaign has been found to vary from yearly and monthly down to daily time scales. 
During a strong VHE flare in November 12-13, 2012 IC~310 has shown an exceptional variability behaviour 
with evidence for extreme short-term variability on (flux doubling) timescales of $\simeq 5$ min, see 
Fig.~\ref{IC310} \citep{2014Sci...346.1080A}.
Jet orientation (probably $i\sim 10-20^o$), power and timing constrains have been taken to disfavour jet-star 
interaction \citep{2012ApJ...749..119B} or magnetic reconnection \citep{2013MNRAS.431..355G} scenarios 
proposed for blazars as the cause of this VHE variability, and the fact that the VHE flux varies on timescales much 
shorter than the light travel time across black hole horizon scales $r_g(3\times10^8 M_{\odot})/c = 25$ min has 
been interpreted as evidence for the occurrence of gap-type particle acceleration on sub-horizon scales (i.e. with 
a gap height $h < $ gravitational scale $ r_g$), see e.g. Fig.~\ref{gap} for illustration. The potential of such particle 
acceleration in vacuum gaps has been explored recently under a variety of conditions, see e.g. 
\citep{2009mfca.book.....B,2011ApJ...730..123L,2015ApJ...809...97B,2016A&A...593A...8P,2016ApJ...818...50H}. 
The observed VHE doubling timescale would imply gap sizes of the order of $h \sim 0.2 r_g$ in the case of IC~310. 
Given that the observed VHE luminosity appears rather high (apparent isotropic VHE luminosity $L_{\rm VHE} \sim 
10^{44}$ erg/s) and that the VHE spectrum is smooth with no indication for absorption up to $8.3$ TeV, gap-type particle 
acceleration scenarios for IC~310 may encounter some challenges as well. Promising work \citep{2016ApJ...818...50H} 
has been started to positively probe its feasibility and further observational characterisation (including the relation of IC~310 
to the BL Lac class) may help to disentangle the nature of the non-thermal VHE processes in IC~310. 
The angular resolution of VHE instruments is moderate only, so that in the absence of variability (e.g. see the case of
 Cen~A) the location of the emitting region is often rather poorly constrained. This if different to IC~310 where the 
exceptional VHE variability now strongly limits possible scenarios and suggests that it may be magnetospheric in origin. 
Given its distance and black hole mass, VLBI-type radio observation will not be able to probe deeply into the near-black 
hole environment of IC~310 (in contrast to the case of M87, see below), so that some uncertainties still remain. We note
 however, that radio observations appear consistent with the ejection of a new component at the time of the TeV activity 
 \citep{Glawion16}.

\subsubsection{M87}
The Virgo Cluster galaxy {\it M87} at a distance of $d\simeq 16.7$ Mpc was the first radio galaxy detected at TeV 
energies \cite{2003A&A...403L...1A}. Its proximity has made it a prime target to probe jet formation scenarios with
high-resolution radio observations down to scales of tens of Schwarzschild radii  $R_s$ (approaching the event 
horizon), and much effort has been recently directed into this, e.g. \citep{2012Sci...338..355D,2016arXiv160805063M,
2015ApJ...803...30K,2015ApJ...807..150A,2016ApJ...817..131H}.
M87 is classified as an FR I source and known to host one of most massive black holes with $M_{\rm BH} \simeq
(2-6)\times 10^9\,M_{\odot}$. At radio frequencies its sub-parsec scale jet appears complex, possibly consisting of 
a slower, mildly relativistic ($\beta \sim 0.5$c) layer and a faster moving spine ($\gamma_b \sim 2.5$), e.g. 
\citep{2016arXiv160805063M}, and to be misaligned by an angle $i \sim (15-25)^{\circ}$. The indications for a
parabolic jet shape may indicate that the jet initially experiences some external (e.g., disk wind) confinement 
\citep{2016MNRAS.461.2605G}. The inferred jet seeds and inclinations are on average consistent with rather 
modest Doppler boosting $D\lppr$ a few (for review, see e.g. \citep{2012MPLA...2730030R}). 
At VHE energies, M87 is known for its particularly interesting characteristics, including rapid day-scale variability 
(on flux doubling time scales $\Delta t_{\rm obs} \sim 1$ d) detected during active source states, and a relatively
hard spectrum compatible with a power law of photons index $\Gamma_{\rm VHE} \simeq 2.2$ and extending from 
$\sim 300$ GeV to beyond $10$ TeV, e.g. \citep{2006Sci...314.1424A,2008ApJ...685L..23A,2012ApJ...746..141A,
2012ApJ...746..151A}. 
Both this hard VHE spectrum and the observed rapid variability are remarkable features for a misaligned AGN, and 
reminiscent of those seen in IC~310.
At HE ($>100$ MeV) gamma-ray energies, Fermi-LAT has reported the detection of gamma-rays form M87 up to 
30 GeV based on 10 months of data with a photon index comparable to the VHE one ($\Gamma_{\rm HE}=2.26
\pm0.13$)\citep{2009ApJ...707...55A} and no indications for variability down to timescales of $\sim 10$ d. A simple 
power-law extrapolation from the HE to the VHE regime would however under-predict the (non-simultaneous) VHE 
fluxes measured during TeV high states, suggesting that the high states might be accompanied by the emergence
of a additional component beyond the conventional assumed SSC-type component. Similar to the case of Cen~A, 
numerous models have been proposed to account for this, and the interested reader is directed to 
Refs.~\citep{2012MPLA...2730030R,2012AIPC.1505...80R} for an overview and discussion of them.\\
The observed rapid VHE variations in M87 are (still) the fastest compared to those seen at any other waveband.
Light travel time arguments point to a compact $\gamma$-ray emitting region ($R < c\Delta t_{\rm obs} D$) with 
a size comparable to the Schwarzschild radius $r_s=(0.6-1.8) \times10^{15}$ cm of its black hole.\\ 
Over the years M87 has shown several active states, with VHE high states being reported for 2005, 2008 and 
2010, and an elevated (TeV flux level 2-3 times higher than average) state detected in 2012. During all VHE high 
states, day-scale VHE variability has been seen. For the elevated state in 2012 evidence for VHE variability on 
timescale of weeks have been reported, hinting to the possibility that the "quiescent" state might perhaps also 
show some longterm evolution \citep{2012AIPC.1505..586B}. High-resolution VLBI radio observations, probing
scales down to tens of gravitational radii, which have been performed during the active 2008, 2010 and 2012 
VHE states provide evidence that the TeV emission is accompanied by (delayed) radio core flux enhancements,
and support the conclusion that the VHE emission may originate at the jet base very near to the black hole 
\cite{2009Sci...325..444A,2012ApJ...760...52H,2014ApJ...788..165H,2015ApJ...807..150A}. The required 
compactness of the TeV zone and the noted radio ($R_s$-scale) -- VHE correlation would seem to support 
models where the observed variable VHE emission is related to gap-type processes occurring in the vicinity 
of the black hole and signals a (fresh) injection of plasma particle in the jet that could lead to some increased 
(delayed, and possibly more extended) radio emission \citep{2011ApJ...730..123L,2015ApJ...809...97B,
2016A&A...593A...8P}. To some extent M87 may thus remind one of IC~310, though the radio core inferences 
seem to make the situation for M87 much more conclusive.

\section{Conclusions}
In recent years non-blazar AGN have emerged as a particularly interesting gamma-ray emitting source class 
on the extragalactic sky. The detection of Narrow Line Seyfert 1 galaxies at Fermi-LAT energies along with 
related multi-wavelength observations is providing important information for our astrophysical understanding 
of jet formation and the trigger mechanisms behind radio-loudness. Radio galaxies, on the other hand, continue 
to significantly impacting on the field. With their jets misaligned and associated Doppler boosting effects only 
modest, they allow unique insights into otherwise "hidden" environments (e.g. close to the black hole, or in 
the "background"). Evidence for an unexpected spectral hardening in the gamma-ray regime for Cen~A, for 
example, are most naturally interpreted as indications for the emergence of a new physical component beyond 
the conventional SSC-type one with possible interpretations ranging from dark matter scenarios to black hole
magnetospheric processes. Extreme VHE variability on timescales smaller than or comparable to the light travel 
time across the horizon of the black hole, as seen for example in IC~310 and M87, on the other hand, suggest 
that this $\gamma-$ray emission may originate in a very compact zone, probably in the vicinity of the black hole 
itself. Further studies in this regard could allow to probe deeply into the Physics of the Extremes. Non-blazar 
AGN, and misaligned sources in particular, are thus turning out to be of increasing astrophysical significance by 
allowing a deep fundamental diagnostic of the non-thermal acceleration and radiation mechanisms, of the link 
between accretion and jet formation processes, of the near black-hole environment and multi-zone emission 
models, thereby deepening our understanding of the nature of cosmic reality. All this makes them sources with 
a very promising physics potential for the upcoming CTA array, both what concerns their spectral as well as their 
timing characterization.

\section{ACKNOWLEDGMENTS}
Financial support by a DFG Heisenberg fellowship (RI 1187/4-1 ) is gratefully acknowledged. 


\nocite{*}
\bibliographystyle{aipnum-cp}%
\bibliography{paper}%

\end{document}